# BIO-INSPIRED METAHEURISTIC OPTIMIZATION FOR HIERARCHICAL ARCHITECTURE DESIGN OF INDUSTRIAL CONTROL SYSTEMS

## *R.M. Zakirzyanov[1]*

[1] NEXT Engineering, LLC
 62, Sibgata Khakima st., Kazan, 421001, Russian Federation

***Abstract.*** *Automated process control systems (APCS) are widely used in modern industrial enterprises. They address three key objectives: ensuring the required quality of manufactured products, ensuring process safety for people and the environment, and reducing capital and operating costs. At large industrial enterprises, APCSs are typically geographically distributed and characterized by a large number of monitored parameters. Such systems often consist of several subsystems built using various technical means and serving different functional purposes. APCSs usually have a hierarchical structure consisting of several levels, where each level hosts commercially available technical devices with predetermined characteristics. This article examines the engineering problem of selecting an optimal software and hardware structure for a distributed process control system applied to a continuous process in the chemical industry. A formal formulation of the optimization problem is presented, in which the hierarchical structure of the system is represented as an acyclic graph. Optimization criteria and constraints are defined. A solution method based on a metaheuristic ant colony optimization algorithm, widely used for this class of problems, is proposed. A brief overview of the developed software tool used to solve a number of numerical examples is provided. The experimental results are discussed, along with parameter selection and possible algorithm modifications aimed at improving solution quality. Information on the verification of the control system implemented using the selected software and hardware structure is presented, and directions for further research are outlined.*

## Introduction

Modern industrial processes are characterized by a high level of automation. In the chemical, petrochemical, and oil refining industries, production is typically carried out in an automatic or automated mode, where the process operator does not directly control the process but instead performs supervisory functions and issues control actions in accordance with technological regulations.

At such enterprises, the automated process control system (APCS) is implemented as a hardware-software system composed of commercially available components [1]. If the controlled object is geographically distributed and involves a large number of variables, the corresponding control system is referred to as a distributed control system (DCS) [2].

The properties and functionality of distributed control systems are largely determined by the system architecture, since the characteristics of the components are predefined by manufacturers and cannot be modified. Therefore, achieving the required system performance necessitates constructing a configuration of components with known characteristics that minimizes cost while satisfying all specified requirements and constraints, i.e., an optimal architecture [3].

The APCS architecture is hierarchical (Fig. 1) and consists of multiple levels.

The number of levels is determined by system designers based on the scale, distribution, and functional requirements of the system. Traditionally, three functional levels of an APCS are distinguished (lower, middle, and upper) [4]. However, in practical implementations, each functional level may consist of several physical layers [5]. For example, the middle level of a local control system may be implemented as a single physical layer based on a programmable logic controller (PLC), whereas the middle level of a APCS may include three or more physical layers (input/output modules, control units, network switches, etc.).

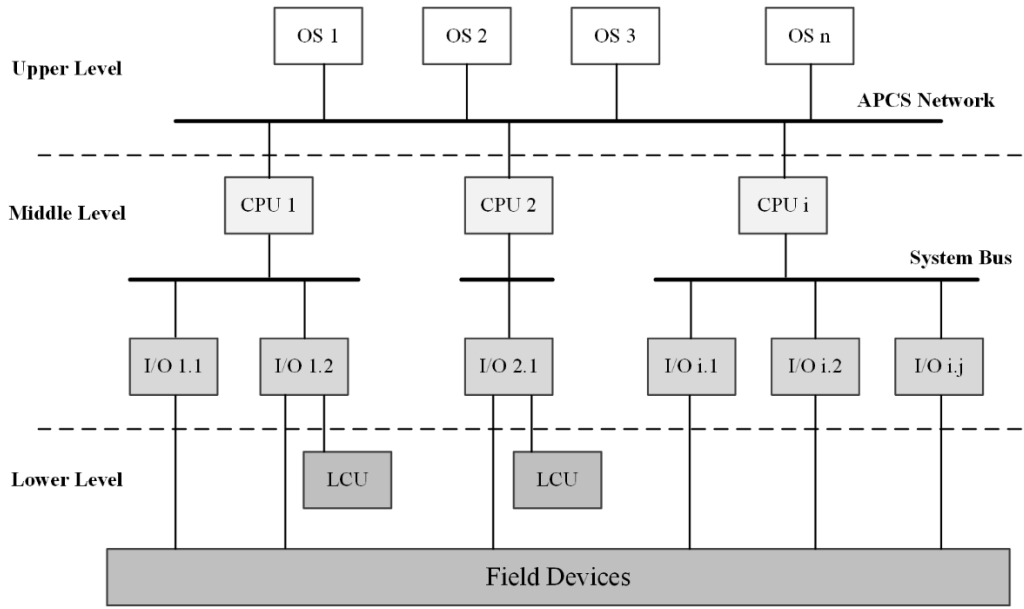

Fig. 1. Typical architecture of an APCS:
CPU – control units (programmable controllers), I/O – input/output modules, OS – operator workstations, LCU – local control units.

In practice, the hardware configuration of an automated process control system (APCS) is determined at an early stage of system design (often before a mathematical model of the controlled object is available) based on initial information about the system (number of signals, reliability, accuracy, and performance requirements). The selected hardware configuration and its characteristics must subsequently be taken into account when developing the mathematical model of the system.

At this stage, the designer must select system components and construct an architecture that ensures effective control of the object during subsequent implementation, while satisfying reliability and performance requirements and minimizing cost.

Designing an optimal hardware architecture for an APCS is an important task that is currently addressed primarily empirically, based on the designer's experience and recommendations from equipment manufacturers. This paper presents a formal formulation of the problem of optimizing the hardware architecture of an APCS, proposes a solution method based on the ant colony optimization (ACO) metaheuristic, and outlines directions for further research in this area.

**Related work**

The problem of optimizing the architecture of technical systems remains highly relevant and continues to attract significant research attention. Various aspects of structural optimization have been addressed in the literature. For example, the optimization of robotic assembly systems is studied in [6], while the construction of optimal architectures for distributed control systems is considered in [7, 8]. Related problems include the optimization of production line structures [9], topology design [10], and chemical process system optimization [11]. The selection of optimal architectures for automated process control systems is discussed in [12], whereas the synthesis of complex system architectures is addressed in [13]. Methods and algorithms for optimizing hierarchical architectures are comprehensively reviewed in [14, 15].

A wide range of solution methods, both exact and approximate, has been developed for such optimization problems. Exact methods are typically computationally expensive, which limits their applicability to large-scale instances. In contrast, approximate methods do not guarantee optimality but can provide near-optimal solutions with moderate computational effort.

Metaheuristic algorithms inspired by biological behavior (bio-inspired algorithms) or natural and social phenomena have become widely used for solving both continuous and discrete optimization problems [16]. A large number of such algorithms have been proposed, including the Genetic Algorithm (GA), Tabu Search (TS), Bat Algorithm (BA), Artificial Bee Colony (ABC), Grey Wolf Optimizer (GWO), Ant Colony Optimization (ACO), Particle Swarm Optimization (PSO), Cuckoo Search (CS), and Simulated Annealing (SA) [17].

Metaheuristic algorithms can be broadly classified into two categories: local search methods, which iteratively improve a single solution (e.g., TS, SA), and population-based methods. The latter include evolutionary algorithms (e.g., GA) and swarm intelligence algorithms (e.g., ACO, PSO, GWO) [18]. A detailed overview of the historical development of metaheuristics is provided in [19].

Metaheuristics have been successfully applied to a wide range of engineering problems [20]. Standard benchmark problems, such as Tension/Compression Spring Design (TCSD), Pressure Vessel Design (PVD), and Welded Beam Design (WBD), are commonly used to evaluate optimization algorithms, along with classical test functions such as Rosenbrock, Rastrigin, and Ackley functions [21].

The Ant Colony Optimization (ACO) algorithm was first proposed by M. Dorigo for solving the Traveling Salesman Problem (TSP) in 1992 [22]. Numerous modifications and extensions of ACO have since been developed [23]. Its applications include controller tuning in control systems [24], route optimization [25], resource planning [26], and facility location problems [27].

Recent research has also focused on hybrid approaches that combine different metaheuristic algorithms [28], as well as the integration of swarm-based methods with local search techniques [29]. In addition, there is growing interest in combining metaheuristics with machine learning methods [30]. Several studies address the problem of parameter selection and tuning for metaheuristic algorithms [31], while others explore automated design of problem-specific evolutionary algorithms [32].

The specific problem addressed in this paper was originally formulated in [33], and the corresponding optimization criteria for distributed control system architectures are presented in [34].

**Problem Statement**

To formulate the problem of finding the optimal hardware architecture of an automated process control system (APCS), the system is represented as a tree (i.e., an acyclic graph), an example of which is shown in Fig. 2. The structure illustrated in the figure may represent, for example, the hardware architecture of the middle functional level of an APCS, which in turn consists of three physical layers.

At Level 3 (the lowest level in the figure), input/output modules are located, to which field devices are directly connected. Level 2 contains controllers that implement the control logic of the process. Level 1 consists of network switches that interconnect multiple controllers into a unified system. The number of levels in the architecture may vary.

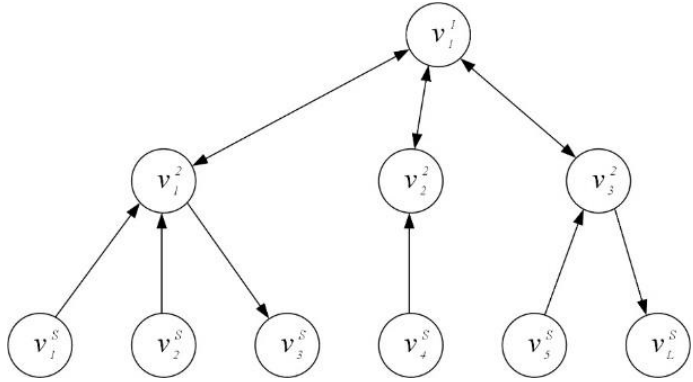

Fig. 2. Hierarchical architecture of an APCS

Let the hierarchical architecture of an automated process control system (APCS) be represented as a graph $\mathcal{G} = (\mathcal{V}, \mathcal{E})$, where $v \in \mathcal{V}$ denotes a device (graph node), and $\mathcal{E}$ is the set of edges representing communication channels between devices.

Let $\mathcal{U} = \{u_1, \dots, u_U\}$ be the set of device types.

It is assumed that each node of the architecture (i.e., each device), regardless of its type, operates according to a cyclic process consisting of the following three phases:

- acquisition of information (reading) from the controlled process or from nodes of the lower hierarchical level;
- processing of information (execution of control algorithms);
- transmission of information (writing) to a lower hierarchical level or generation of control actions applied to the process.

Each device in the system is assigned one of two roles: a processing unit or a relay unit. A processing unit executes control algorithms (e.g., controllers or servers), whereas a relay unit performs only data transmission functions (e.g., network switches or input/output modules).

Each device type $u_i \in \mathcal{U}$ is characterized by the following parameters:

- $C_i \in \mathbb{R}^+$ – device cost;
- $N_i \in \mathbb{N}$ – number of physical channels;
- $R_i \in \mathbb{R}^+$ – maximum memory capacity;
- $P_i \in [0,1]$ – device failure probability;
- $T_i \in \mathbb{R}^+$ – execution time of a single instruction (performance);
- $y_i \in \{0,1\}$ – operating mode (1 – processing unit, 2 – relay unit);
- $M_i \in \mathbb{N}$ – maximum number of child devices;

- $\tau_i \in \mathbb{R}^+$ – transmission delay for relay units (for processing units $\tau_i = 0$).

Each device type is represented by the vector defined in (1).

$$u_i = (C_i, N_i, R_i, P_i, T_i, y_i, M_i, \tau_i)^T, \quad i = \overline{1, U}. \tag{1}$$

Let $\mathcal{A} = \{a_1, \dots, a_A\}$ be the set of control loops, where each loop $a_j \in \mathcal{A}$ is defined as:

$$a_j = (n_j, r_j, w_j)^T, \quad j = \overline{1, A}, \tag{2}$$

where $n_j \in \mathbb{N}$ is the number of physical signals in the loop, $r_j \in \mathbb{R}^+$ is the memory required to store all instructions and variables of the loop, and $w_j \in \mathbb{N}$ is the number of instructions in the loop processing program, characterizing its computational complexity.

In general, a loop may contain different types of signals (analog, discrete, input, output, etc.). In this study, it is assumed that all field signals are pre-assigned to loops, and that for each loop the number of signals, the number of instructions in the processing program, and the required memory are known (or estimated). It is further assumed that the execution time of any instruction is constant and equal to $T_i$.

The optimal control system architecture is represented as a tree with $S \in \mathbb{N}$ levels, where $s = 1$ corresponds to the root (top of the hierarchy), and $s = S$ corresponds to the leaf level (bottom of the hierarchy). Let $K_s$ denote the number of devices at level $s$, and let $V \in \mathbb{N}$ be the total number of devices in the system.

Let $L \in \mathbb{N}$ denote the number of leaf nodes (devices directly connected to field equipment), and let $\mathcal{L} \subseteq \mathcal{V}$ denote the set of leaf nodes. Then, $\mathcal{L} = \{v_1^S, \dots, v_L^S\}$.

Based on the above, for each device $v_k^s \in \mathcal{V}$, the following mappings are introduced:

- $u(v_k^s) = u_k^s \in \mathcal{U}$ – device type;
- $\pi(v_k^s) \in \mathcal{V} \cup \{\emptyset\}$ – parent node;
- $\mathcal{D}(v_k^s) \subseteq \mathcal{V}$ – set of child nodes;
- $\mathcal{D}'(v_k^s) \subseteq \mathcal{V}$ – subtree rooted at node $v_k^s$, where $\mathcal{D}(v_k^s) \subseteq \mathcal{D}'(v_k^s)$;
- $\mathcal{P}(v_k^s) \subseteq \mathcal{V}$ – set of nodes on the path from $v_k^s$ to the root, including the root.

Here, $s \in \{1, \dots, S\}$ denotes the hierarchy level, and $k \in \mathbb{N}$ is the index of the node at level $s$. Thus, each device can be represented as the vector defined in (3).

$$v_k^s = \left(u_k^s, \pi(v_k^s), \mathcal{D}(v_k^s), \mathcal{D}'(v_k^s)\right)^T, \quad s = \overline{1, S}, \quad k = \overline{1, K_s}. \tag{3}$$

To solve the problem, it is assumed that only devices at level $S$ (leaf nodes) are directly connected to field equipment of the process plant, i.e., they receive measurements from the process and apply control actions.

It is further assumed that all devices belong to a predefined set of types and differ only in the values of the parameters introduced above. Horizontal connections between devices are not allowed.

Each control loop must be physically connected to a leaf node of the tree. A leaf node $v$ can process the assigned loops directly if it is a processing unit ($y = 1$), or forward them to an upper-level processing unit if it is a relay unit ($y = 0$). Each loop must be assigned to exactly one processing unit. A processing unit can process only those loops that are connected to leaf nodes within its subtree.

It is assumed that only relay units introduce transmission delays. Processing units execute assigned loops cyclically. The execution time of a processing unit is equal to the total processing time of all loops assigned to it.

To simplify the notation, indices $s$, $k$, and $j$ are omitted below.

- $x_{va} \in \{0,1\}$, where $x_{va} = 1$, if control loop $a$ is physically connected to leaf $v$ ($\forall v \in \mathcal{L}, \forall a \in \mathcal{A}$), and $x_{va} = 0$ otherwise;
- $z_{va} \in \{0,1\}$, where $z_{va} = 1$, if control loop $a$ is assigned to processing unit $v$ ($\forall v \in \mathcal{V}, \forall a \in \mathcal{A}$), and $z_{va} = 0$ otherwise.

Thus, if a control loop is physically connected to a node at level $S$, it is considered to be connected to a leaf node $v$ (i.e., $x_{va} = 1$). If the signals of loop $a$ are processed at node $v$, then loop $a$ is assigned to node $v$ (i.e., $z_{va} = 1$).

The following constraints define the tree structure of the distributed control system.

The tree has a single root (4). The root node may be either a processing unit or a relay unit.

$$\exists!\, v_1^1 \in \mathcal{V}: \pi(v_1^1) = \emptyset. \tag{4}$$

Nodes located at level $S$ (leaf nodes) do not have child nodes:

$$|\mathcal{D}(v)| = 0, \quad \forall v \in \mathcal{L}. \tag{5}$$

Each node can have at most $M_i$ child devices:

$$|\mathcal{D}(v)| \leq M_i, \quad \forall v \in \mathcal{V} \backslash \mathcal{L}, \tag{6}$$

where $u(v) = u_i$ denotes the type of node $v$. Constraint (6) applies to all nodes except leaves.

Each control loop must be connected to exactly one leaf node:

$$\sum_{v \in \mathcal{L}} x_{va} = 1, \quad \forall a \in \mathcal{A}. \tag{7}$$

The following constraints define the assignment of control loops to processing units. Only processing units can process control loops; therefore, a loop cannot be assigned to a relay unit:

$$z_{va} \leq y_v, \quad \forall v \in \mathcal{V}, \quad \forall a \in \mathcal{A}. \tag{8}$$

Each control loop must be assigned to exactly one processing unit:

$$\sum_{v \in \mathcal{V}} z_{va} = 1, \quad \forall a \in \mathcal{A}. \tag{9}$$

Each processing unit is subject to a memory constraint. The total memory required by all assigned loops must not exceed the available memory:

$$\sum_{a \in \mathcal{A}} z_{va} r_a \leq R_v, \quad \forall v \in \mathcal{V}. \tag{10}$$

A processing unit can process only those loops that are connected to leaf nodes within its subtree. Thus, a processing unit serves only its own subtree:

$$z_{va} \leq \sum_{\substack{v' \in \mathcal{L} \\ v' \in \mathcal{D}'(v)}} x_{v'a}, \quad \forall v \in \mathcal{V}, \quad \forall a \in \mathcal{A}. \tag{11}$$

The total number of signals from all control loops connected to a leaf node $v$ must not exceed the number of available channels:

$$\sum_{a \in \mathcal{A}} x_{va}\, n_a \leq N_v, \quad \forall v \in \mathcal{V}. \tag{12}$$

Here, $N_v$ denotes the number of channels determined by the device type of node $v$. A node may have $N_v = 0$.

Since the designed architecture corresponds to a control system for a dynamic process, it is important to take into account its dynamic properties during synthesis. In practice, this is a complex problem [13].

As a performance indicator, the maximum processing time of a control loop is considered. For any loop $a$, the total processing time $T_a$ must not exceed a specified limit $T_{max} \in \mathbb{R}^+$.

The processing time of a control loop consists of two components: (i) the execution time of all instructions corresponding to loops assigned to the processing unit, and (ii) the total transmission delay introduced by relay units along the path from the leaf node to the processing unit and back (13).

$$T_a = \max_{a \in \mathcal{A}} \begin{bmatrix} \sum_{a' \in \mathcal{A}} z_{v(a)a'} \cdot w_{a'} \cdot T_{v(a)} + \\ +2 \cdot \sum_{v' \in \mathcal{P}(v(a),a)} (1 - y_{v'}) \tau_{v'} \end{bmatrix} \leq T_{max}, \qquad (13)$$

where, $v(a)$ denotes the processing unit assigned to control loop $a$, and $\mathcal{P}(v(a), a)$ denotes the path from the leaf node to which loop $a$ is connected to the processing unit $v(a)$. The term $(1 - y_{v'})$ is an indicator function equal to 1 if node $v'$ is a relay unit and 0 if it is a processing unit.

When designing the system, reliability must be taken into account. The probability of failure is used as a reliability indicator. It is assumed that the system becomes inoperable if at least one of its components fails, and that failures of individual devices are independent.

Under these assumptions, the system failure probability is given by

$$P_{sys} = 1 - \prod_{v \in \mathcal{V}} (1 - P_v) \leq P_{max}. \qquad (14)$$

where $P_v$ is the failure probability determined by the type of device $v$.

In this study, the failure probability is assumed to be constant. In real systems, it is generally time-dependent, i.e., $P_v = f(t)$ [36]. Methods for reliability evaluation of hierarchical automated process control systems are described in [35, 36].

The following constraints define the coordination of processing units. A processing unit cannot have a parent that is also a processing unit:

$$y_v + y_{\pi(v)} \leq 1, \quad \forall v \neq v_1^1. \qquad (15)$$

A processing unit cannot have other processing units in its subtree:

$$y_v + \sum_{v' \in \mathcal{D}'(v)} y_{v'} = 1, \quad \forall v \in \mathcal{V} : y_v = 1. \qquad (16)$$

There must be exactly one processing unit on the path from any leaf node to the root:

$$\sum_{v' \in \mathcal{P}(v)} y_{v'} = 1, \quad \forall v \in \mathcal{L}, \qquad (17)$$

Previously, key criteria for optimizing the hierarchical architecture of a distributed control system were identified in [34].

The total cost is a primary factor in system design. In general, system cost consists of capital and operating costs. Assuming that the cost of each device is fixed, the total implementation cost is defined as the sum of the costs of all devices.

The optimal hierarchical architecture $\mathcal{G}_O$ is obtained by minimizing the total system cost under constraints (4)-(17):

$$C_O = \min_{\mathcal{G}} \sum_{v \in \mathcal{V}} C_v, \qquad (18)$$

where $C_O$ is the optimal total cost of the system.

Thus, the structural optimization problem can be formulated as follows: determine a hierarchical architecture composed of devices of predefined types that minimizes the objective function (18) subject to constraints (4)-(17).

It is assumed that an unlimited number of devices of each type is available. The resulting architecture is assumed to be a homogeneous tree, i.e., the path from any leaf node to the root contains the same number of levels.

**Method Selection and Software Implementation**

The problem described above is a combinatorial (discrete) optimization problem with multiple constraints. This problem is NP-hard because, under certain assumptions, it can be reduced to a known NP-complete problem (e.g., the knapsack problem [37]).

Exact methods (e.g., dynamic programming [6]) or approximate methods (e.g., metaheuristic algorithms) can be used to solve the problem. Exact methods are computationally expensive and difficult to scale, and therefore may be ineffective for large problem instances. Therefore, heuristic and metaheuristic algorithms are widely used to solve such problems.

Selecting a suitable metaheuristic algorithm is a nontrivial task in itself [38]. To select an appropriate solution method, an analysis of the performance of common optimization algorithms on typical problems (benchmarks) was conducted [17, 21]. In [17], it was shown that the GWO, BA, and CS algorithms perform best when solving continuous optimization problems, while GA, ABC, and ACO are more suitable for solving discrete problems.

The genetic algorithm (GA) is designed for systems with a fixed structure; therefore, when applied to a tree construction problem, difficulties may arise in encoding solutions. The artificial bee colony algorithm (ABC) does not store intermediate results and is not designed for working with tree structures. The ant colony optimization (ACO) algorithm stores intermediate results, allows for convenient encoding of solutions, as it was originally developed for solving graph problems, is simple to implement, and can be easily combined with other algorithms. Therefore, it was chosen for solving the problem of optimizing the structure of an automated process control system.

A description of the ant colony algorithm is given in [22, 37]. A distinctive feature of the algorithm is that it stores the best solutions using traces (pheromones) that evaporate at a certain rate.

Each step is selected according to the following rule:

$$P_i = \frac{\tau_i^{\alpha} \eta_i^{\beta}}{\sum_{k=0}^{N} \tau_k^{\alpha} \eta_k^{\beta}}, \qquad (19)$$

where $P_i$ is the probability of selecting the $i$-th option, $\tau_i$ is the pheromone level associated with the $i$-th option, $\eta_i$ is the heuristic value characterizing the cost (preference) of the $i$-th option, $\alpha$ is the parameter controlling the influence of pheromone ("herding"), $\beta$ is the parameter controlling the influence of the heuristic ("greed"), and $N$ is the total number of feasible options.

The pheromone level is updated according to the following rule:

$$\tau_{i+1} = (1 - \rho)\tau_i + \Delta\tau_i, \qquad (20)$$

where $\tau_{i+1}$ is the pheromone level at iteration $i + 1$, $\tau_i$ is the pheromone level at iteration $i$, $\rho$ is the pheromone evaporation coefficient, and $\Delta\tau_i$ is the amount of deposited pheromone.

In this problem, each selection of a new device to be added to the tree is performed with a probability determined by expression (19).

The ant incrementally constructs the tree from the root to the leaves by adding graph nodes. The construction process is completed when a feasible tree is obtained. After construction, all constraints are verified. If the constructed tree satisfies all constraints, it is considered a valid solution; otherwise, it is discarded.

In simplified form, the algorithm consists of several main steps, described below. Each selection of a new device is performed by the ACO algorithm according to the rules defined by expressions (19) and (20). The flowchart of the algorithm is shown in Fig. 3. The steps in which the ACO algorithm selects a new device are highlighted in color.

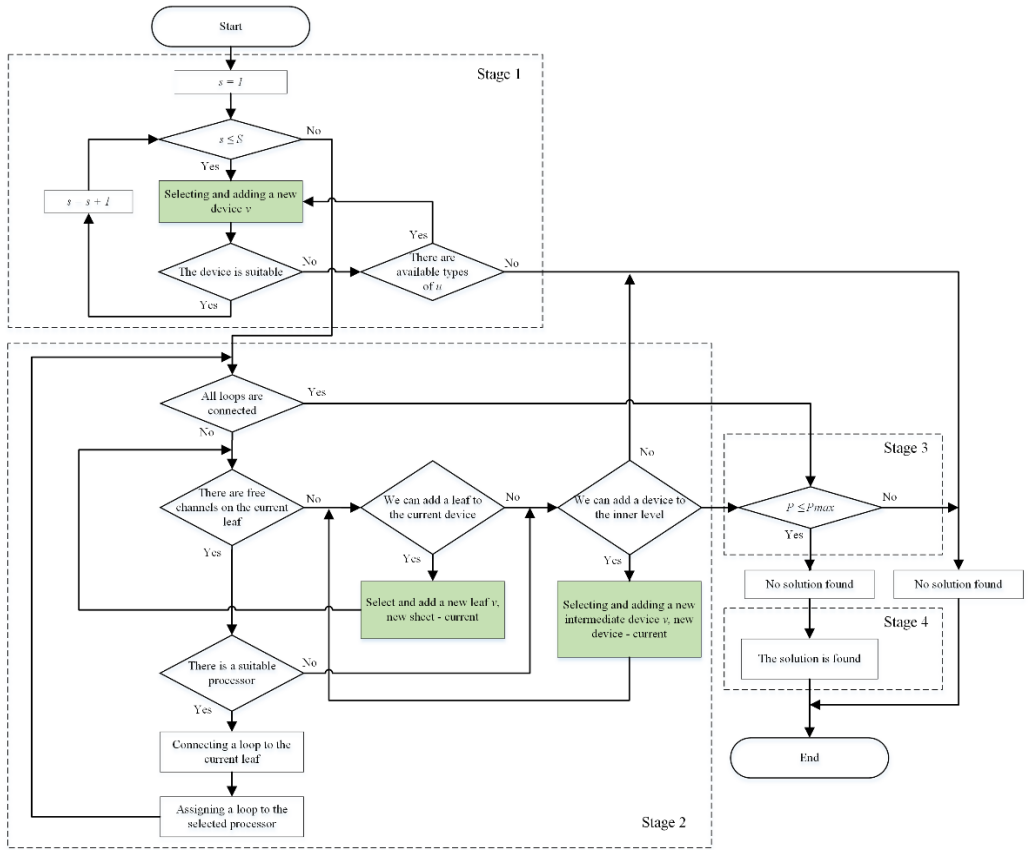

Fig. 3. Flowchart of the algorithm for constructing the optimal hardware architecture

*Stage 1. Construct a tree backbone of height S, with the process unit level defining.* Select an arbitrary device (either a processing unit or a relay unit) and place it at the top level of the hierarchy.

If the root is a processing unit, only relay units can be added as its children, since there must be exactly one processing unit on any path from a leaf to the root. If the root is a relay unit, both processing units and relay units can be added as its children.

Select the next device and place it at the next lower level of the hierarchy. If a processing unit has already been placed above, only a relay unit can be selected. If no processing unit has yet been placed in the backbone, either a processing unit or a relay unit can be selected.

Repeat this procedure until level $S$ (the leaf level) is reached. At the leaf level, only devices that support connection of field signals (control loops) can be used, i.e., devices with $N_i > 0$. The level at which the first processing unit is placed is defined as the processing unit level. All subsequent processing units can be placed only at this level.

*Stage 2. Connecting Control Loops by Adding Leaves and Branches.* We begin by connecting control loops to the leaf nodes (initially, there is only one leaf). The first control loop is selected and connected to the leaf. Each loop requires a certain amount of memory, $r_i$, for processing. We traverse the tree upward until a processing unit is found. This processing unit is assigned to process the loop.

We then check whether the selected processing unit can process the loop. The available memory of the processing unit must be greater than or equal to the total memory required by all loops assigned to it. If there is insufficient memory, the processing unit is replaced with another one, and the check is repeated. If no suitable processing unit can be found, the solution does not exist.

If the processing unit has sufficient memory, the processing time constraint is verified. The sum of transmission delays $\tau_i$ introduced by all relay units along the path from the leaf node (to which the loop is connected) to the assigned processing unit and back, plus the processing time of all loops assigned to this processing unit, must not exceed the specified value $T_{max}$.

If there are still available channels on the leaf node, the next control loop is connected using a greedy assignment strategy [39]. If no channels are available, it is checked whether an additional device can be added at this level of the hierarchy. If the parent node allows adding another child device, a new device is selected and added at the leaf level.

If no additional child device can be added (i.e., the maximum number of child devices $M_i$ has been reached), higher levels of the hierarchy are examined. At each level, it is checked whether a new child device can be added. When such a possibility exists, a new device is selected and added to the hierarchy.

If the new device is added at the leaf level, the connection of control loops continues. If the device is added at a higher level, the corresponding branch is extended to the leaf level. If it is not possible to extend the branch to the leaf level (e.g., due to selecting a device that does not support child connections), the algorithm backtracks to the upper level and changes the device type until a feasible option is found.

If channels are available on a leaf node, a control loop is connected, feasibility is verified, and the loop is assigned to a processing unit.

The procedure continues until all control loops are connected and assigned to processing units.

*Stage 3. Constraint Checking.* Once all control loops are connected and assigned to processing units, the system reliability is evaluated using formula (14). If the system failure probability exceeds the specified value $P_{max}$, the solution is discarded. It should be noted that constraints (4)-(13) and (15)-(17) are satisfied by construction and do not require additional verification.

*Stage 4. Objective Function Calculation.* The system cost objective function is evaluated using formula (18). The cost is defined as the sum of the costs of all system components (i.e., all devices in the hierarchy). The goal of the algorithm is to find a feasible tree with minimal cost.

After constructing a feasible solution, the pheromone values are updated according to formula (20), and the next ant is initialized. The number of ants in the population is one of the key tuning parameters of the ACO algorithm.

The algorithm is executed over multiple iterations. At each iteration, the best feasible solution is retained. The number of iterations is also configurable. To implement the algorithm, a Python program [40] was developed, capable of solving numerical problems of relatively large scale.

The software generates a tree structure that satisfies the problem constraints. Its functionality allows for the integration of additional solution methods (through the implementation of corresponding modules), visualization of the resulting system architecture, and generation of convergence plots. The convergence plot can be exported in *.png or *.csv format. The main interface of the program is shown in Fig. 4.

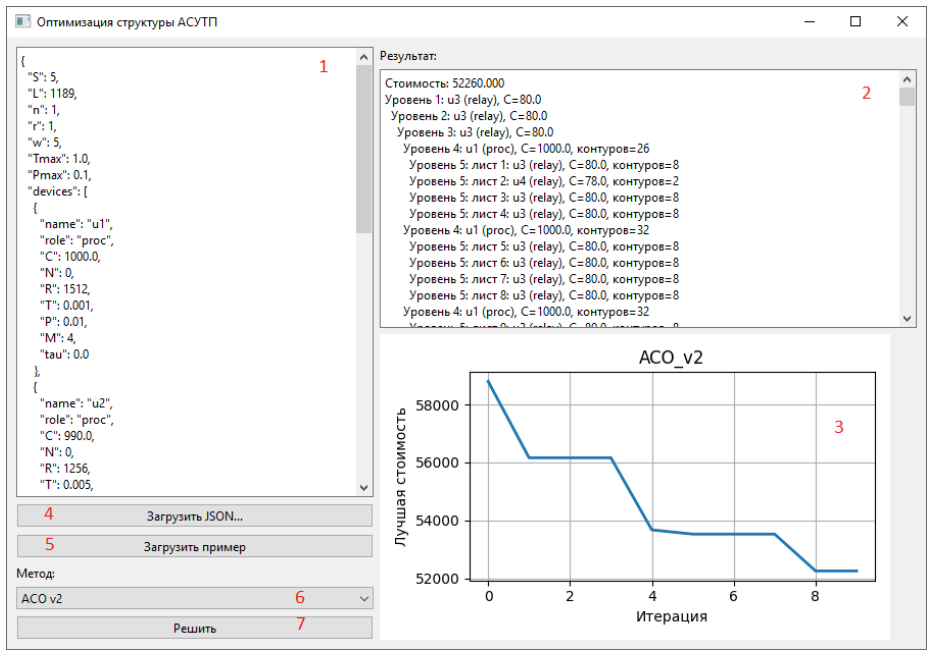

Fig. 4. Main window interface of the software:
1 – input field for problem data in JSON format; 2 – output field displaying the resulting system architecture as a tree; 3 – convergence plot of the algorithm; 4 – button for loading input data from a JSON file; 5 – button for loading a default example of input data; 6 – dropdown list for selecting the solution method; 7 – button for running the algorithm.

The program accepts the following input data in JSON format: the number of hierarchy levels, the number and types of control loops, constraint limits, a description of available device types (processing units and relay units), and algorithm parameters ($\alpha$, $\beta$, $\rho$, etc.). The program runs on Windows and Linux operating systems.

**Experimental Results**

Using the developed software, a number of numerical examples were solved, and the influence of algorithm parameters on the solution search process was studied. Testing was performed on a workstation with the following specifications: Intel Core i5-7400 CPU (3.0 GHz), 16 GB RAM, 512 GB SSD, and Windows 10 Pro operating system.

The input data for the numerical example are presented in Table 1. The example considers a set of \( A \) identical control loops with parameters $n_1 = 1$, $r_1 = 1$, $w_1 =$

5. The maximum processing time of a control loop, $T_{max} = 1$ s, and the maximum allowable system failure probability, $P_{max} = 0,1$, were used as constraints.

**Input data for the numerical example**

| Device type $u_i$ | $u_1$ | $u_2$ | $u_3$ | $u_4$ | $u_5$ |
|---|---|---|---|---|---|
| Cost $C_i$, arb. units | 1000 | 990 | 80 | 78 | 65 |
| Number of channels $N_i$, units | 0 | 0 | 8 | 4 | 2 |
| Memory capacity $R_i$, units | 512 | 256 | - | - | - |
| Failure probability $P_i$ | 0,01 | 0,02 | 0,005 | 0,005 | 0,001 |
| Instruction execution time $T_i$, s | 0,002 | 0,004 | - | - | - |
| Processing unit flag $y_i$ | 1 | 1 | 0 | 0 | 0 |
| Max. number of child nodes $M_i$, units | 4 | 4 | 4 | 4 | 8 |
| Transmission delay $\tau_i$, s | 0 | 0 | 0,01 | 0,03 | 0,02 |

Processing units $u_1$ and $u_2$ may represent, for example, programmable logic controllers (PLCs), which are relatively expensive but do not include I/O channels. Relay units $u_3$, $u_4$, and $u_5$ may represent I/O modules, to which 2 to 8 physical signals can be connected, or network switches. All device types are equipped with network ports (4 to 8) for connecting downstream devices.

The algorithm was executed with a population of 20 ants and 20 iterations. The following ACO parameters were used: $\rho = 0,25$, $\alpha = 2$, $\beta = 1$. For each set of input data, 20 independent runs of the algorithm were performed. The results obtained for different values of $A$ and $S$ are summarized in Table 2.

The following notation is used in Table 2: $A$ is the number of control loops, $S$ is the number of hierarchy levels, $C_{min}$ is the best value of the objective function, $C_{avg}$ is the average value of the objective function, $CV = (\sigma/C_{avg}) \cdot 100\%$ is the coefficient of variation, $W$ is the percentage of successful runs, and $T$ is the average execution time of the algorithm.

**Experimental Results**

| $A$, units | $S$, units | $C_{min}$, arb. units. | $C_{avg}$, arb. units. | $CV$, % | $W$, % | $T$, s |
|---|---|---|---|---|---|---|
| 1 | 3 | 1120 | 1120,00 | 0,00 | 100 | 0,013 |
| 5 | 3 | 1135 | 1135,00 | 0,00 | 100 | 0,019 |
| 10 | 3 | 1200 | 1201,15 | 0,29 | 100 | 0,032 |
| 15 | 3 | 1215 | 1215,00 | 0,00 | 100 | 0,040 |
| 20 | 3 | 1293 | 1293,20 | 0,05 | 100 | 0,045 |
| 25 | 3 | 1360 | 1362,80 | 0,36 | 100 | 0,062 |
| 30 | 3 | 1375 | 1375,00 | 0,00 | 100 | 0,061 |
| 35 | 3 | 1453 | 1454,20 | 0,15 | 100 | 0,082 |
| 40 | 3 | 1455 | 1456,00 | 0,21 | 100 | 0,081 |
| 45 | 3 | 1535 | 1540,50 | 1,27 | 100 | 0,092 |
| 50 | 3 | 1675 | 1688,25 | 0,42 | 100 | 0,102 |
| 55 | 3 | 1690 | 1690,00 | 0,00 | 100 | 0,109 |
| 60 | 3 | 1768 | 1770,90 | 0,27 | 100 | 0,136 |
| 65 | 3 | 1835 | 1846,55 | 0,27 | 100 | 0,132 |
| 70 | 3 | 1850 | 1850,00 | 0,00 | 100 | 0,140 |
| 75 | 3 | 1928 | 1929,20 | 0,05 | 100 | 0,169 |
| 80 | 3 | 1930 | 1932,05 | 0,25 | 100 | 0,161 |

| $A$, units | $S$, units | $C_{min}$, arb. units. | $C_{avg}$, arb. units. | $CV$, % | $W$, % | $T$, s |
|---|---|---|---|---|---|---|
| 85 | 3 | 2010 | 2027,20 | 1,70 | 100 | 0,171 |
| 90 | 3 | 2075 | 2095,15 | 1,19 | 100 | 0,180 |
| 95 | 3 | 2183 | 2185,70 | 0,19 | 100 | 0,216 |
| 100 | 3 | 5065 | 5079,10 | 0,21 | 100 | 0,201 |
| 120 | 3 | 5225 | 5244,90 | 0,15 | 100 | 0,215 |
| 140 | 3 | 6465 | 6486,90 | 0,17 | 100 | 0,224 |
| 160 | 3 | 6635 | 6731,57 | 4,39 | 70 | 0,224 |
| 180 | 3 | - | - | - | 0 | - |
| 180 | 4 | 4741 | 5492,00 | 19,73 | 100 | 0,349 |
| 200 | 4 | 6235 | 8141,65 | 14,97 | 100 | 0,340 |
| 250 | 4 | 7391 | 10420,90 | 9,53 | 100 | 0,370 |
| 300 | 4 | 13221 | 13243,45 | 0,11 | 100 | 0,433 |
| 350 | 4 | 14700 | 14784,85 | 0,26 | 100 | 0,517 |
| 400 | 4 | 17249 | 17288,25 | 0,11 | 100 | 0,545 |
| 450 | 4 | 19816 | 19883,05 | 0,19 | 100 | 0,589 |
| 500 | 4 | 21273 | 21363,15 | 0,18 | 100 | 0,650 |
| 600 | 4 | 25330 | 25591,20 | 1,75 | 100 | 0,666 |
| 700 | 4 | 29350 | 30283,67 | 4,43 | 30 | 0,717 |
| 800 | 4 | - | - | - | 0 | - |
| 800 | 5 | 33570 | 33660,65 | 0,14 | 100 | 0,877 |
| 900 | 5 | 38735 | 38782,15 | 0,07 | 100 | 1,010 |
| 1000 | 5 | 42706 | 42787,70 | 0,12 | 100 | 1,160 |
| 1500 | 5 | 63135 | 63213,70 | 0,08 | 100 | 2,067 |
| 2000 | 5 | 84439 | 84557,40 | 0,06 | 100 | 2,977 |
| 2500 | 5 | 105764 | 108980,50 | 6,33 | 100 | 3,615 |
| 3000 | 5 | 126409 | 148825,60 | 12,75 | 25 | 3,851 |
| 5000 | 5 | - | - | - | 0,00 | - |
| 5000 | 6 | 210896 | 212027,65 | 1,19 | 100 | 8,534 |
| 10000 | 6 | 421875 | 483879,67 | 13,26 | 75 | 25,328 |

Overall, the algorithm demonstrates good performance. It is noteworthy that for $A = 180$ and $S = 3$, the algorithm fails to find feasible solutions, since the given structure does not provide a sufficient number of devices to connect all control loops, and the constraints prevent the addition of new devices. In this case, increasing the number of hierarchy levels is required to ensure connectivity, as shown in the table.

It should also be noted that as the tree approaches its maximum capacity, the solution quality deteriorates significantly. This is due to the fact that the feasible solution space becomes highly restricted, and the algorithm is unable to find an initial feasible solution within the specified number of iterations. Possible approaches to address this issue include increasing the number of iterations (or the population size) or modifying the algorithm, for example by incorporating local search.

The convergence plots of the algorithm for selected values of $A$ and $S$ are shown in Fig. 5. The horizontal axis represents the iteration number, and the vertical axis represents the objective function (system cost).

A graphical representation of the resulting hierarchical architectures for selected values of $A$ and $S$ is shown in Fig. 6. The numbers in the figure indicate the number of connected control loops (for leaf nodes) and assigned control loops (for processing units). Each device type is represented by a distinct color (a "warm" color scheme is used for processing units, and a "cold" color scheme is used for relay units).

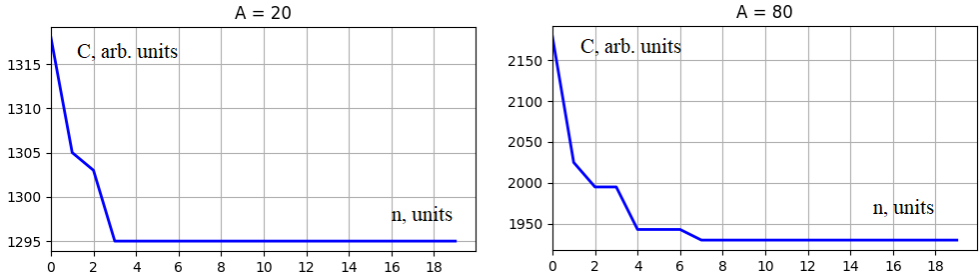

Fig. 5. Convergence plots of the algorithm for A = 20 and A = 80 for S = 3

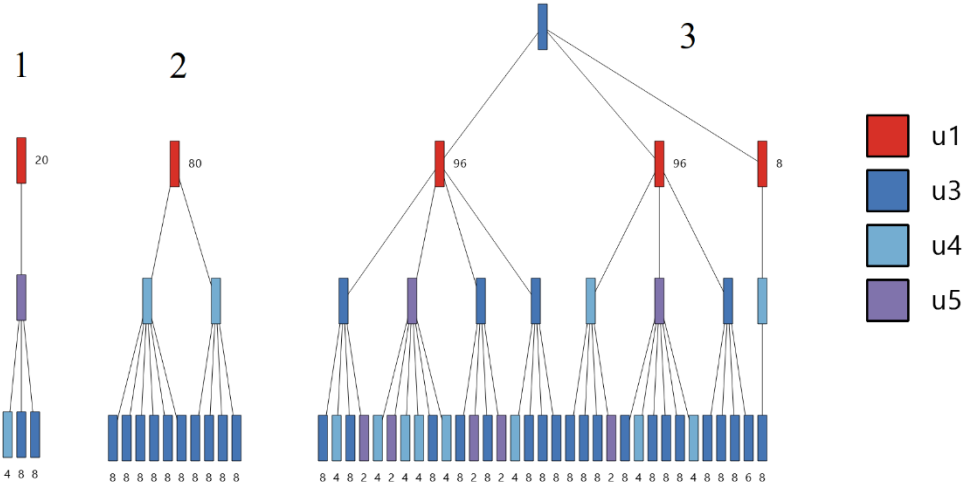

Fig. 6. System architectures for different values of *A* and *S*:
1 – A = 20 and S = 3; 2 – A = 80 and S = 3; 3 – A = 200 and S = 4

### Modification and Selection of Algorithm Parameters

The ant colony optimization (ACO) algorithm is an approximate method and does not guarantee an exact solution to the problem. However, it can be used to obtain a baseline solution, which may subsequently be improved [41].

Like many metaheuristic algorithms, ACO is sensitive to the choice of tuning parameters. The key parameters of the algorithm are the pheromone influence coefficient $\alpha$, the heuristic influence coefficient $\beta$, and the pheromone evaporation coefficient $\rho$. When $\alpha = 0$ and $\beta = 0$, the algorithm reduces to random search. Fig. 7 shows the convergence plots of the algorithm for one dataset ($A = 500$) from Table 2. The results demonstrate that ACO significantly outperforms random search.

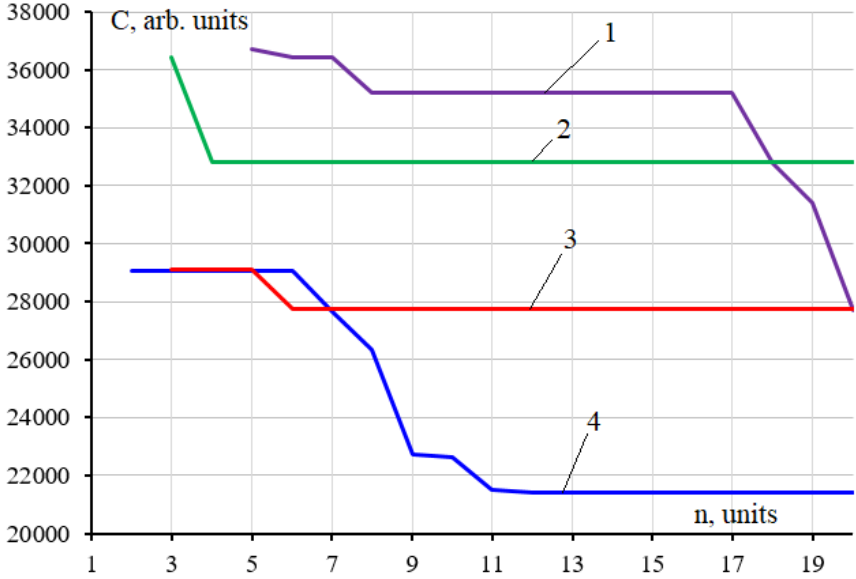

Fig. 7. Convergence plots of the algorithm for different values of $\alpha$ and $\beta$:
$1 - \alpha = 0$ and $\beta = 1$; $2 - \alpha = 0$ and $\beta = 0$; $3 - \alpha = 2$ and $\beta = 0$; $4 - \alpha = 2$ and $\beta = 1$

Numerical experiments show that the problem contains regions with either too many or too few feasible solutions. In such regions, the algorithm either fails to find a solution within a given number of iterations or converges to the optimum very slowly. Since, in these regions, preference should be given to devices with a large number of channels, a promising approach is to modify the heuristic used in the algorithm (see formula (19)) to increase the attractiveness of such devices.

In practice [36], the most commonly used heuristic characterizes the cost of a step (in this case, the selected device):

$$\eta = \frac{1}{c_i}, \tag{21}$$

where $C_i$ is the cost of the $i$-th device.

To improve the algorithm's ability to find feasible solutions, it is proposed to modify the heuristic to favor low-cost devices with a large number of channels:

$$\eta = \frac{N_i}{c_i}, \tag{22}$$

where $N_i$ is the number of channels (i.e., the maximum number of connected control loops) of the $i$-th device.

Promising directions for further research include fine-tuning of algorithm parameters and the development of more complex heuristics (including adaptive ones) to improve performance. Furthermore, it should be noted that the ACO algorithm in its pure form is rarely used for solving engineering problems [23, 29]. To avoid premature convergence to local optima, it is typically combined with local search methods or other metaheuristic algorithms (e.g., GA or ABC). If the initial tree is constructed using ACO, it is reasonable to consider applying a genetic algorithm to further improve the solution, since it operates on a fixed structure for which it is well suited.

**Input data for the design of the IMS architecture**

| Device type $u_i$ | $u_1$ | $u_2$ | $u_3$ | $u_4$ | $u_5$ | $u_6$ | $u_7$ |
|---|---|---|---|---|---|---|---|
| Device model | PLC210-01-CS | NLScon-A40-S | IGS-500T | IGS-800T | MB210-101 | NLS-8AI-Ethernet | NLS-16AI-I-Ethernet |
| Price (including VAT) $C_i$, RUB | 77544 | 91988 | 8968 | 13285 | 22977 | 28182 | 21350 |
| Number of channels $N_i$, units | 0 | 0 | 0 | 0 | 8 | 8 | 16 |
| Memory capacity $R_i$, MB | 256 | 1024 | - | - | - | - | - |
| Failure probability $P_i$ | 0,0839 | 0,0839 | 0,0839 | 0,0839 | 0,0839 | 0,0839 | 0,0839 |
| AI Instruction execution time $T_i$, s | $10^{-6}$ | $0{,}8 \times \times 10^{-6}$ | - | - | - | - | - |
| Processing unit flag $y_i$ | 1 | 1 | 0 | 0 | 0 | 0 | 0 |
| Max. number of child nodes $M_i$, units | 1 | 1 | 5 | 8 | 0 | 0 | 0 |
| Transmission delay $\tau_i$, s | 0 | 0 | 0 | 0 | 0,6 | 0,28 | 0,08 |

**Verification**

The proposed algorithm was applied to the design of the hardware architecture of the information and measurement system (IMS) for the TGME-428 power boiler. The design engineers were required to develop the hierarchical architecture of the IMS at an early stage, given incomplete information about the controlled object.

During the development of the IMS, equipment from OWEN (PLC210) [42], RealLab (NLScon PLC) [43], and Planet (network switches) [44] was considered, as presented in Table 3. The parameter values were determined based on the manufacturers' documentation. Prices were obtained from the official websites of the manufacturers as of February 24, 2026.

The design was carried out for an IMS containing 260 analog signals (for each signal: $n_1 = 1$, $r_1 = 2 \cdot 10^{-4}$, $w_1 = 1$. The resulting optimal system architecture is shown in Fig. 8.

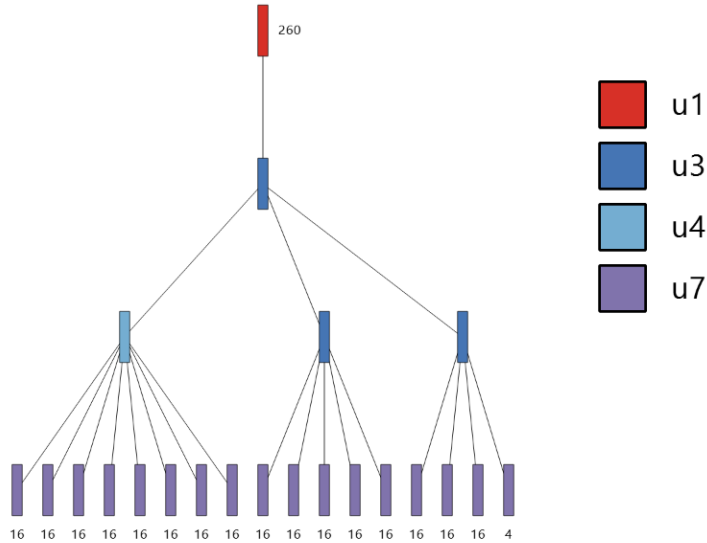

Fig. 8. Hardware architecture of the IMS for the TGME-428 boiler:
$u_1$ – PLC210; $u_3$ – network switch IGS-500T; $u_4$ – network switch IGS-800T; $u_7$ – analog input module NLS-16AI-I-Ethernet

The minimum system cost is 480,683 RUB. The minimum number of hierarchy levels is $S = 4$.

**Conclusion**

The problem of finding an optimal hardware architecture for an automated process control system (APCS) based on commercially available components is highly relevant. At early design stages (e.g., feasibility studies), system developers are required to determine the system architecture and estimate its cost under conditions of limited information about the controlled object. In practice, such architectures are typically selected empirically, based on the designer's experience and manufacturers' recommendations, and are not guaranteed to be optimal.

The results presented in this paper enable the construction of a reference near-optimal APCS architecture under multiple constraints. This architecture, which defines communication links between components as well as node performance and load, should be taken into account when developing mathematical models of control systems.

The proposed algorithm can be used by design engineers and APCS developers for the design of large-scale, geographically distributed control systems, where communication delays, device performance, and computational load are critical factors.

Future research directions include: comparison of the proposed ACO-based approach with other metaheuristic and exact methods; fine-tuning of algorithm parameters and analysis of their impact on solution quality; development of hybrid approaches combining ACO with local search and other metaheuristics (e.g., GA, ABC, GWO) to improve convergence and avoid local optima; and investigation of machine learning techniques for solving the problem.In addition, further studies may focus on the influence of constraints on the solution process and on refining the formal problem formulation by relaxing or revising the adopted assumptions.